\documentclass[10pt,twocolumn,secnumarabic,amssymb, nobibnotes, aps, prd,superscriptaddress]{revtex4-2}
\usepackage{xcolor,soul}
\sethlcolor{red}
\usepackage{amsmath}
\usepackage{wrapfig}
\usepackage[utf8]{inputenc}  
\usepackage{braket}

\usepackage{lipsum}
\usepackage{subfigure}
\usepackage{wrapfig}
\usepackage{xparse,xcoffins}
\usepackage{calligra}
\DeclareMathAlphabet{\mathcalligra}{T1}{calligra}{m}{n}
\DeclareFontShape{T1}{calligra}{m}{n}{<->s*[2.2]callig15}{}

\ExplSyntaxOn
\NewCoffin\imagecoffin
\NewCoffin\labelcoffin

\keys_define:nn { miguel/label }
 {
  label   .tl_set:N = \l_miguel_label_tl,
  labelbox .bool_set:N = \l_miguel_label_box_bool,
  labelbox .default:n = true,
  fontsize .tl_set:N = \l_miguel_label_size_tl,
  fontsize .initial:n = \footnotesize,
  pos .choice:,
  pos/nw .code:n = \tl_set:Nn \l_miguel_label_pos_tl { left,up },
  pos/ne .code:n = \tl_set:Nn \l_miguel_label_pos_tl { right,up },
  pos/sw .code:n = \tl_set:Nn \l_miguel_label_pos_tl { left,down },
  pos/se .code:n = \tl_set:Nn \l_miguel_label_pos_tl { right,down },
  pos/n .code:n = \tl_set:Nn \l_miguel_label_pos_tl { hc,up },
  pos/w .code:n = \tl_set:Nn \l_miguel_label_pos_tl { left,vc },
  pos/s .code:n = \tl_set:Nn \l_miguel_label_pos_tl { hc,down },
  pos/e .code:n = \tl_set:Nn \l_miguel_label_pos_tl { right,vc },
  pos .initial:n = nw,
  unknown .code:n   = \clist_put_right:Nx \l_miguel_label_clist
                       { \l_keys_key_tl = \exp_not:n { #1 } }
 }
\clist_new:N \l_miguel_label_clist
\box_new:N \l_miguel_label_box
\box_new:N \l_miguel_label_image_box

\NewDocumentCommand{\xincludegraphics}{O{}m}
 {
  \group_begin:
  \tl_clear:N \l_miguel_label_tl
  \clist_clear:N \l_miguel_label_clist
  \keys_set:nn { miguel/label } { #1 }
  \tl_if_empty:NTF \l_miguel_label_tl
   {
    \miguel_includegraphics:Vn \l_miguel_label_clist { #2 }
   }
   {
    \SetHorizontalCoffin\imagecoffin
     {
      \miguel_includegraphics:Vn \l_miguel_label_clist { #2 }
     }
    \SetHorizontalCoffin\labelcoffin
     {
      \raisebox{\depth}
       {
        \bool_if:NTF \l_miguel_label_box_bool
         { \fcolorbox{white}{white}{\l_miguel_label_size_tl\l_miguel_label_tl} }
         { \l_miguel_label_size_tl\l_miguel_label_tl }
       }
     }
    \SetVerticalPole\imagecoffin{left}{3pt+\CoffinWidth\labelcoffin/2}
    \SetVerticalPole\imagecoffin{right}{\Width-3pt-\CoffinWidth\labelcoffin/2}
    \SetHorizontalPole\imagecoffin{up}{\Height-3pt-\CoffinHeight\labelcoffin/2}
    \SetHorizontalPole\imagecoffin{down}{3pt+\CoffinHeight\labelcoffin/2}
    \use:x{\JoinCoffins\imagecoffin[\l_miguel_label_pos_tl]\labelcoffin[vc,hc]}
    \TypesetCoffin\imagecoffin
   }
   \group_end:
 }
\NewDocumentCommand{\setlabel}{m}
 {
  \keys_set:nn { miguel/label } { #1 }
 }

\cs_new_protected:Nn \miguel_includegraphics:nn
 {
  \includegraphics[#1]{#2}
 }
\cs_generate_variant:Nn \miguel_includegraphics:nn { V }

\ExplSyntaxOff

\makeatletter
\newcommand{\twocolumncaption}{\@dblarg\@twocolumncaption}
\def\@twocolumncaption[#1]#2{%
  \renewcommand{\@makecaption}[2]{%
    \par\vskip\abovecaptionskip\begingroup\small\rmfamily
    \splittopskip=0pt
    \setbox\@tempboxa=\vbox{
      \@arrayparboxrestore \let \\\@normalcr
      \hsize=.5\hsize \advance\hsize-1em
      \let\\\heading@cr
      \noindent ##1\ ##2\par
    }%
    \vbadness=10000
    \setbox\z@=\vsplit\@tempboxa to .55\ht\@tempboxa
    \setbox\z@=\vtop{\hrule height 0pt \unvbox\z@}
    \setbox\tw@=\vtop{\hrule height 0pt \unvbox\@tempboxa}
    \noindent\box\z@\hfill\box\tw@\par
    \endgroup\vskip \belowcaptionskip
  }%
  \setlength{\abovecaptionskip}{4ex}%
  \caption[#1]{#2}%
}

\usepackage{graphicx}
\begin{document}
\title{First- and Second-Order Digital Quantum Simulation of Three-Level Jaynes--Cummings Dynamics on Superconducting Quantum Processors
}

\author{J. Thirunirai Selvam}
\author{S. Saravana Veni}
    \email{s_saravanaveni@cb.amrita.edu }
    \affiliation{Department of Physics, Amrita School of Physical Sciences,
 Amrita Vishwa Vidyapeetham, Coimbatore, 641112, Tamil Nadu, India
}
\author{Ria Rushin Joseph}
\email{ria.joseph@deakin.edu.au}
\affiliation{School of IT, Deakin University, Melbourne, Australia}
\begin{abstract}
This work presents a digital quantum simulation of a three-level atomic system
interacting with a single-mode electromagnetic field based on the Jaynes--Cummings
model, implemented on IBM Quantum superconducting processors. A qutrit is encoded
using two physical qubits to represent the atomic states, while an additional qubit
encodes the truncated field mode, enabling the realization of effective
$\Lambda$-type atomic dynamics.The continuous-time light--matter interaction is implemented in a digital form by
discretizing the evolution using Suzuki--Trotter decomposition. In contrast to an
analog realization, the digital simulation replaces the continuous evolution with a
sequence of quantum gates whose parameters are explicitly controlled. Phase
evolution arising from the interaction Hamiltonian is digitally encoded using
calibrated $R_Z$ gates, whose rotation angles are fixed by the physically relevant
coupling scale and the chosen Trotter time step.State preparation is achieved using Hadamard and parametrized rotation gates, while
the interaction dynamics are implemented through controlled operations. A
comparative analysis between first- and second-order Trotter implementations
reveals a trade-off between digital accuracy and hardware-induced noise. Overall,
the results demonstrate that calibrated gate operations and noise-aware circuit
design enable reliable digital simulation of multi-level light--matter interactions
on noisy intermediate-scale quantum platforms.
\end{abstract}
\maketitle
\section{Introduction}

Quantum computing exploits the principles of quantum mechanics to perform
computational tasks that are intractable for classical computers\cite{Alexeev2021}. Quantum information processing is fundamentally based on
quantum bits (qubits), which, unlike classical bits, can exist in coherent
superpositions of states and exhibit entanglement. These uniquely quantum
features enable powerful computational paradigms that offer exponential or
polynomial speedups for problems such as integer factorization, optimization,
and the simulation of quantum many-body systems\cite{GillBuyya2026,Doyle2026}.

Among the various physical realizations of qubits, superconducting circuits
have emerged as a leading platform due to their scalability, compatibility
with established semiconductor fabrication techniques, and fast, high-fidelity
gate operations \cite{AbuGhanem2026,ArmaghaniRostami2026}. Superconducting qubits are engineered from
nonlinear, non-dissipative electrical circuits incorporating Josephson
junctions, which give rise to discrete, quantized energy levels. Operated at
cryogenic temperatures, typically below $20~\mathrm{mK}$, these circuits
behave as artificial atoms with controllable parameters \cite{Zhang2019,Frolov2013,AwschalomLoss2002,Gambetta2017,Kwon2021}.

The transmon qubit is the most widely used superconducting qubit architecture\cite{Larsen2026,Lange2026}.
It is a modified charge qubit designed to suppress sensitivity to charge noise
by increasing the ratio of Josephson energy to charging energy
. The resulting weak anharmonicity allows selective control of
the lowest two energy levels, which define the computational basis states
$\ket{0}$ and $\ket{1}$. Using tunable couplings and precisely shaped microwave
pulses, superconducting qubits can be entangled and manipulated to implement
universal quantum logic operations\cite{Peterer2015,Roth2021}. Large-scale quantum
processors developed by companies such as IBM, Google, and Rigetti now
incorporate tens to hundreds of superconducting qubits, enabling experimental
studies of near-term quantum algorithms and device-level physics\cite{Bravyi2022,Bayerstadler2021}.

Despite this progress, current quantum processors operate in the Noisy
Intermediate-Scale Quantum (NISQ) regime, characterized by limited coherence
times and the absence of fully fault-tolerant quantum error correction. Quantum states are highly susceptible to both intrinsic and
environmental noise sources, including material defects, control imperfections,
and high-energy radiation events. The mitigation of such noise through improved
materials, device design, and error-correcting strategies remains a central
challenge in superconducting quantum technologies\cite{RempferObenland2024,Zhang2026Continual}.

While the conventional Jaynes--Cummings (JC) model describing the interaction
between a two-level atom and a single quantized field mode has been studied
extensively, increasing attention has been directed toward multi-level atomic
systems owing to their richer internal structure and enhanced control
capabilities\cite{MedinaDozal2026,Casanova2010,Smirne2010,Cius2025}. In particular, three-level atomic configurations such as ladder,
$\Lambda$-, and V-type systems introduce quantum interference effects that can
significantly modify population transfer, coherence properties, and
atom--field entanglement dynamics. These features are not only of fundamental
interest in quantum optics but are also directly relevant to quantum information
processing, quantum control, and state engineering protocols\cite{McCollumMirza2026,Torosov2015,Bina2010}.

Existing studies on three-level Jaynes--Cummings-type models have largely
focused on analytical treatments and idealized numerical simulations,
highlighting phenomena such as modified Rabi oscillations, population trapping,
and interference-induced suppression or enhancement of entanglement. However,
most of these investigations assume noise-free dynamics and do not explicitly
address the constraints imposed by contemporary quantum hardware. As a result,
the practical feasibility of realizing multi-level light--matter interaction
models on present-day quantum processors remains an open and important issue\cite{sanchez2026jaynes,liang2025dynamical}.The rapid development of superconducting quantum processors has opened new avenues for exploring complex quantum dynamical systems beyond the capabilities of classical computation. Among such systems, the Jaynes--Cummings model occupies a central position in quantum optics, cavity quantum electrodynamics, and quantum information science, as it provides a fundamental description of light--matter interaction at the single-quantum level. While the ideal Jaynes--Cummings dynamics are well understood theoretically, extending these models to multi-level atoms and experimentally relevant regimes introduces significant computational and experimental \cite{chong2025long,wu2026efficient,wei2024n,li2024exploring}.

Classical numerical simulations of multi-level atom--field interactions suffer from an exponential growth of the Hilbert space with increasing excitation number and atomic complexity, limiting their applicability to small systems and short evolution times. At the same time, analog quantum simulators based on cavity or circuit QED platforms require precise hardware engineering and offer limited tunability and scalability. Digital quantum simulation using programmable superconducting qubits provides a flexible alternative, enabling systematic control over Hamiltonian parameters, interaction strengths, and evolution times through gate-based implementations\cite{marinkovic2022complexity,kurman2026powering,sharmila2020tomographic}.

In the noisy intermediate-scale quantum (NISQ) era, however, practical quantum simulations are constrained by decoherence, gate imperfections, and readout errors. These limitations impose strict bounds on circuit depth and simulation accuracy, particularly for time-dependent and interacting quantum systems\cite{koch2020demonstrating,guimaraes2024optimized}. Therefore, it is essential to develop resource-efficient circuit constructions and benchmarking strategies that balance physical fidelity with experimental feasibility. In this context, the comparison between first- and second-order Suzuki--Trotter decompositions offers valuable insights into the trade-off between digital accuracy and hardware noise\cite{suzuki1986quantum,ostmeyer2023optimised,jones2019optimising}.

Motivated by these challenges, the present work aims to design, implement, and experimentally evaluate optimized quantum circuits for simulating a three-level Jaynes--Cummings system on real superconducting quantum hardware. By integrating physically motivated parameter calibration, hardware-aware fidelity estimation, and systematic backend selection, this study seeks to establish practical guidelines for reliable quantum optical simulations in near-term devices. The results contribute toward bridging the gap between theoretical quantum optics models and their experimental realization on scalable quantum computing platforms\cite{azuma2011quantum,mischuck2013qudit, moiseev2016quantum,shaw2026cavity}.

\section{Derivation of Atom--Field Dynamics and Entanglement}

The interaction between a two-level atom and a single quantized mode of
the electromagnetic field provides the simplest fully quantum model for
studying light--matter interaction. In cavity and circuit quantum
electrodynamics, this interaction is well described by the
Jaynes--Cummings Hamiltonian under the rotating-wave approximation (RWA),
\begin{equation}
H = \hbar \omega_f a^\dagger a
+ \frac{\hbar \omega_a}{2}\sigma_z
+ \hbar g \left( a\sigma_+ + a^\dagger \sigma_- \right),
\end{equation}
where $\omega_f$ is the frequency of the cavity field mode and
$\omega_a$ is the atomic transition frequency between the excited state
$|e\rangle$ and the ground state $|g\rangle$. The operators $a^\dagger$
and $a$ create and annihilate photons in the cavity and satisfy bosonic
commutation relations. The Pauli operator $\sigma_z$ describes the atomic
population inversion, while $\sigma_+$ and $\sigma_-$ govern atomic
excitation and relaxation processes. The coupling constant $g$ sets the
rate at which energy is coherently exchanged between the atom and the
field and depends on the dipole moment of the atom and the cavity mode
volume.

The rotating-wave approximation is essential in this context because it
eliminates rapidly oscillating counter-rotating terms that average out
on experimentally relevant time scales. This approximation ensures
energy conservation during atom--field interaction and allows for an
exact analytical treatment of the dynamics, which is crucial for
understanding entanglement generation mechanisms.

Under the resonance condition $\omega_f=\omega_a$, the Hamiltonian
commutes with the total excitation number operator
\begin{equation}
\hat{N} = a^\dagger a + \sigma_+\sigma_-,
\end{equation}
such that
\begin{equation}
[H,\hat{N}] = 0.
\end{equation}
This conservation law is not merely a mathematical convenience; it
reflects the physical fact that excitations are exchanged coherently
between the atom and the field without loss. As a result, the full
Hilbert space decomposes into invariant subspaces of fixed excitation
number. Each subspace is spanned by the states
$\{|e,n\rangle, |g,n+1\rangle\}$, which share the same total excitation.
This restriction dramatically simplifies the dynamics and makes it
possible to derive exact time-dependent solutions\cite{larson2021jaynes}.

Assuming an initial atom--field product state with the atom in its
excited state and the field prepared in a Fock state,
\begin{equation}
|\psi(0)\rangle = |e,n\rangle,
\end{equation}
the system evolution remains confined to the corresponding excitation
subspace. Consequently, the time-dependent state can be expressed as
\begin{equation}
|\psi(t)\rangle = c_e(t)|e,n\rangle + c_g(t)|g,n+1\rangle,
\end{equation}
where the probability amplitudes $c_e(t)$ and $c_g(t)$ encode the
exchange of excitation between atom and field.

Substituting this state into the time-dependent Schrödinger equation
\begin{equation}
i\hbar \frac{d}{dt}|\psi(t)\rangle = H|\psi(t)\rangle
\end{equation}
and projecting onto the basis states yields the coupled equations
\begin{align}
i\dot{c}_e(t) &= g\sqrt{n+1}\,c_g(t), \\
i\dot{c}_g(t) &= g\sqrt{n+1}\,c_e(t).
\end{align}
The appearance of the factor $\sqrt{n+1}$ originates from the bosonic
nature of the field and highlights the photon-number dependence of the
atom--field coupling strength. This dependence is responsible for
nonclassical effects such as collapse and revival phenomena when the
field is initially in a superposition of Fock states.

Solving the coupled equations with initial conditions
$c_e(0)=1$ and $c_g(0)=0$ gives
\begin{align}
c_e(t) &= \cos\!\left(g\sqrt{n+1}\,t\right), \\
c_g(t) &= -i\sin\!\left(g\sqrt{n+1}\,t\right),
\end{align}
leading to the time-evolved state
\begin{equation}
|\psi(t)\rangle =
\cos\!\left(g\sqrt{n+1}\,t\right)|e,n\rangle
- i\sin\!\left(g\sqrt{n+1}\,t\right)|g,n+1\rangle.
\end{equation}
This solution demonstrates coherent Rabi oscillations between atomic and
field excitations with a photon-number-dependent Rabi frequency
\begin{equation}
\Omega_n = 2g\sqrt{n+1}.
\end{equation}
Such oscillations are a hallmark of strong coupling and have been
observed in cavity and circuit QED experiments\cite{liu2024quantum}.

To quantify quantum correlations generated during the interaction, the
total atom--field density operator is defined as
\begin{equation}
\rho_{AF}(t) = |\psi(t)\rangle\langle\psi(t)|.
\end{equation}
Although the composite system remains in a pure state, the atomic
subsystem alone generally does not. Tracing over the field degrees of
freedom,
\begin{equation}
\rho_A(t) = \mathrm{Tr}_F\!\left[\rho_{AF}(t)\right],
\end{equation}
yields the reduced atomic density matrix
\begin{equation}
\rho_A(t) =
\begin{pmatrix}
\cos^2\!\left(g\sqrt{n+1}\,t\right) & 0 \\
0 & \sin^2\!\left(g\sqrt{n+1}\,t\right)
\end{pmatrix}.
\end{equation}
The mixed nature of $\rho_A(t)$ directly reflects the entanglement
between the atom and the field, even in the absence of decoherence . 

\begin{figure*}[htbp]
    \centering
    \includegraphics[width=0.5\linewidth]{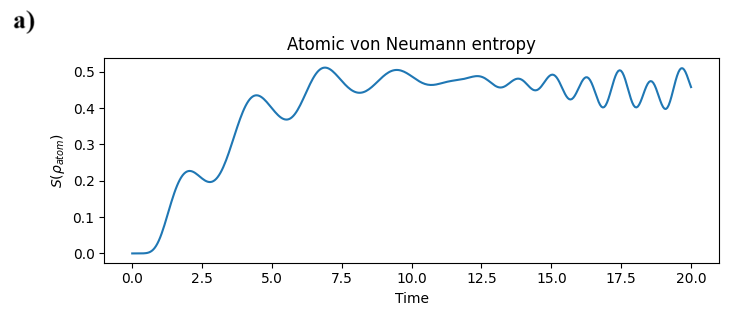} \hfill
    \includegraphics[width=0.48\linewidth]{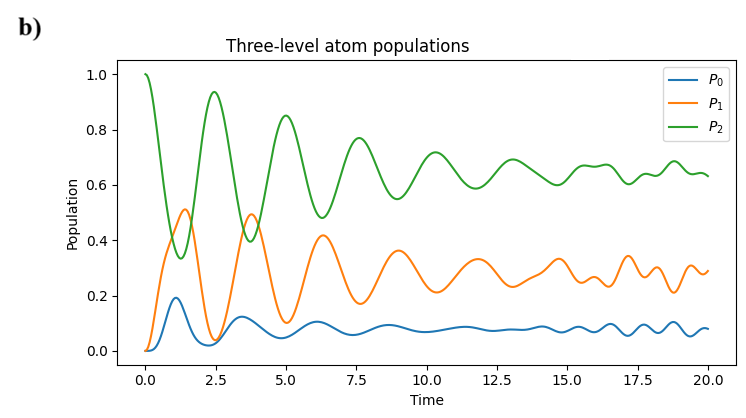}
    \caption{
    \textbf{(a)}  The initial coherent field state, where
the atom is prepared in the excited state and the cavity field is in a
coherent state with mean photon number $\bar{n}$. Since a coherent
state is a superposition of many Fock states, each photon-number
component evolves with a slightly different Rabi frequency. This
results in dephasing among the components, causing the atomic
von Neumann entropy to increase and exhibit oscillatory behavior.
The entropy does not return exactly to zero within the plotted time
range because the revival time is longer than the displayed interval.
No decoherence or dissipation is included in the numerical simulation;
the observed behavior arises purely from coherent, unitary
Jaynes--Cummings dynamics.
    \textbf{(b)} Time evolution of the populations of the three atomic levels under
    anti–Jaynes--Cummings couplings. The populations $P_0$, $P_1$, and $P_2$ exhibit
    coherent oscillations and population redistribution, indicating reversible
    excitation exchange and $\Lambda$-type three-level atomic dynamics.
    }
    \label{fig:entropy_populations}
\end{figure*}

The degree of entanglement is quantified using the atomic von Neumann
entropy
\begin{equation}
S_A(t) = -\mathrm{Tr}\!\left[\rho_A(t)\ln\rho_A(t)\right].
\end{equation}
Starting from zero for an initially separable state, the entropy
increases as atom--field correlations build up andFigure~\ref{fig:entropy_populations} shows the time evolution of the atomic entropy. Starting from
a nearly pure atomic state, the entropy initially increases due to atom--field
entanglement generated by the interaction. The subsequent oscillatory behavior reflects
periodic exchange of quantum information between the atom and the field, analogous to
collapse and revival phenomena observed in Jaynes--Cummings dynamics. The saturation
of entropy at later times indicates the formation of a quasi-steady entangled state\cite{villas2019multiphoton}.

Beyond two-level systems, multilevel atoms enable richer interference
phenomena such as coherent population trapping and electromagnetically
induced transparency. We consider a three-level $\Lambda$-type atom
interacting with a weak probe field and a strong control field.

Under the rotating-wave approximation (RWA), the interaction between the
three-level atom and the applied classical fields is described by the
time-dependent Hamiltonian
\begin{equation}
H_I(t) = -\frac{\hbar}{2}
\left(
g_p(t)\,\sigma_{31}
+
g_c(t)\,\sigma_{32}
+ \text{H.c.}
\right),
\end{equation}
where $\sigma_{31} = |3\rangle\langle 1|$ and $\sigma_{32} = |3\rangle\langle 2|$
are the atomic transition operators associated with the
$|1\rangle \leftrightarrow |3\rangle$ and
$|2\rangle \leftrightarrow |3\rangle$ transitions, respectively.
The quantities $g_p(t)$ and $g_c(t)$ denote the time-dependent coupling strengths
(Rabi frequencies) of the probe and control fields.

The explicit time dependence of $g_p(t)$ and $g_c(t)$ allows for the description
of shaped laser pulses or slowly varying field envelopes, enabling dynamical
control of the atomic population and coherence.
Such temporal modulation plays a crucial role in coherent control protocols,
including electromagnetically induced transparency (EIT), adiabatic population
transfer, and dark-state formation.
The Hermitian conjugate (H.c.) terms ensure that the Hamiltonian remains
self-adjoint and account for the reverse atomic transitions.\cite{fleischhauer2005electromagnetically}.

The Rabi frequencies $\Omega_p$ and $\Omega_c$ characterize the strengths of the
coherent probe and control fields, respectively, and can be independently tuned
experimentally. The resulting dynamics therefore correspond to a coherently driven
$\Lambda$-type three-level atomic system within the rotating-wave approximation (RWA),
rather than an anti--Jaynes--Cummings interaction.

The atomic state is written as
\begin{equation}
|\psi(t)\rangle = c_1(t)|1\rangle + c_2(t)|2\rangle + c_3(t)|3\rangle,
\end{equation}
where $|1\rangle$ and $|2\rangle$ are the two lower states and $|3\rangle$ is the excited
state. Substituting this expansion into the Schrödinger equation with the RWA
interaction Hamiltonian yields the coupled equations of motion
\begin{align}
\dot{c}_1(t) &= i\frac{\Omega_p}{2}c_3(t), \\
\dot{c}_2(t) &= i\frac{\Omega_c}{2}c_3(t), \\
\dot{c}_3(t) &= i\frac{\Omega_p}{2}c_1(t)
+ i\frac{\Omega_c}{2}c_2(t).
\end{align}

These equations explicitly show that the excited state $|3\rangle$ acts as a coherent
intermediate level that couples the two lower states through the applied fields.
Quantum interference between the probe-driven and control-driven excitation pathways
leads to coherent population oscillations and, for suitable parameter regimes,
partial population trapping. This mechanism underlies phenomena such as dark-state
formation and electromagnetically induced transparency (EIT), which have no analogue
in two-level atomic systems~\cite{arimondo1996v}.

The level populations are obtained from the reduced atomic density matrix as
\begin{equation}
P_j(t) = \langle j|\rho_A(t)|j\rangle, \quad j=1,2,3.
\end{equation}

Figure~\ref{fig:entropy_populations} illustrates the temporal evolution of these
populations. The oscillatory transfer of population between the three levels reflects
coherent multi-frequency beating driven by the probe and control fields. The apparent
envelope modulation arises from interference between distinct excitation pathways,
rather than from dissipative decay processes (unless explicit decoherence terms are
included).

Such behavior is a hallmark of coherently driven three-level $\Lambda$ systems and
demonstrates the richer coherence control and population dynamics achievable beyond
the standard two-level Jaynes--Cummings model.

\section{Construction of First- and Second-Order Digital Circuits for a Three-Level Jaynes--Cummings System}

\begin{figure*}[htbp]
\centering
\includegraphics[width=0.82\linewidth]{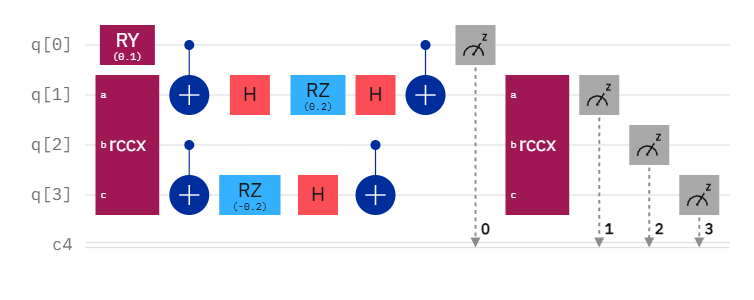}\hfill
\includegraphics[width=0.82\linewidth]{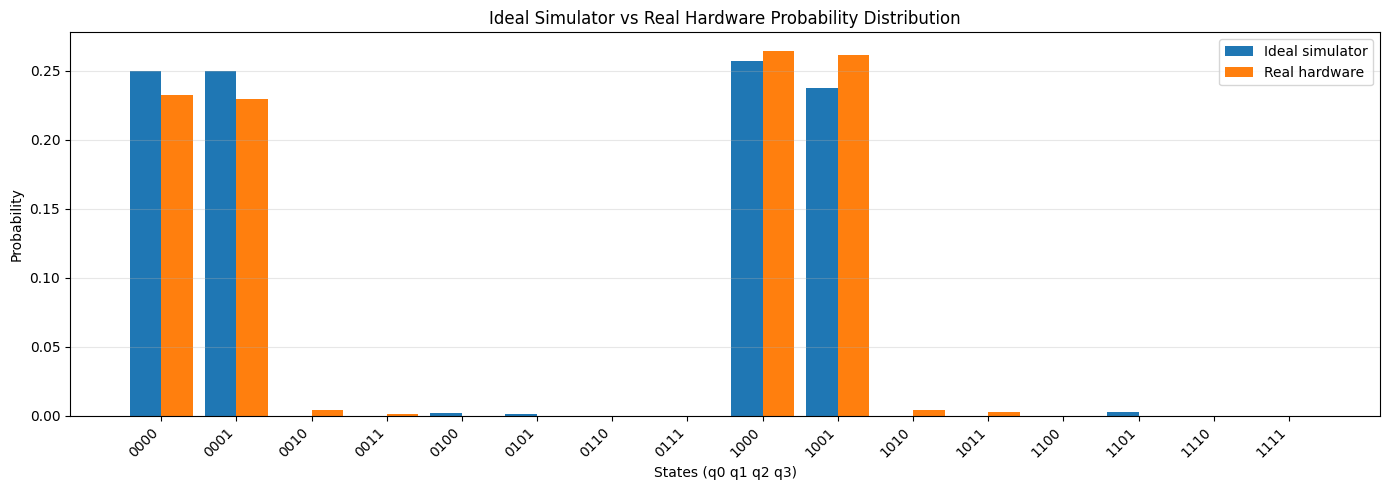}\hfill
\includegraphics[width=0.52\linewidth]{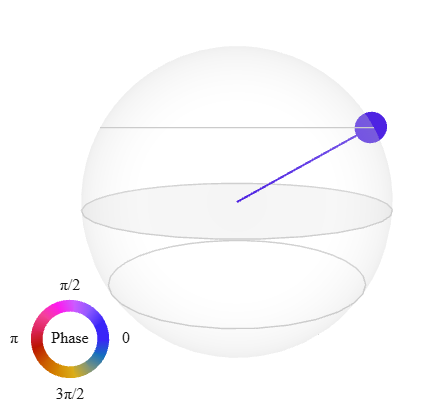}
\caption{
First-order Suzuki--Trotter simulation of the three-level
Jaynes--Cummings model.
\textbf{(a)} Digital quantum circuit implementing the first-order
decomposition of the interaction Hamiltonian.
\textbf{(b)} Measurement probability distribution obtained from
1024 shots on real hardware, showing dominant populations within
allowed excitation subspaces.
In comparison with the corresponding ideal (noise-free) simulator,
small but nonzero populations appear in classically forbidden states,
arising from gate imperfections and decoherence.
\textbf{(c)} Q-sphere plots are generated from simulator-derived statevectors (or reconstructed quantum states), whereas hardware executions directly yield measurement counts in selected bases. Consequently, Q-sphere visualizations should be interpreted as simulation-based diagnostics of phase and amplitude structure, unless full or partial quantum state tomography or an entanglement witness is additionally performed.
}
\label{fig_2}
\end{figure*}

\begin{figure*}[htbp]
\centering
\includegraphics[width=0.8\linewidth]{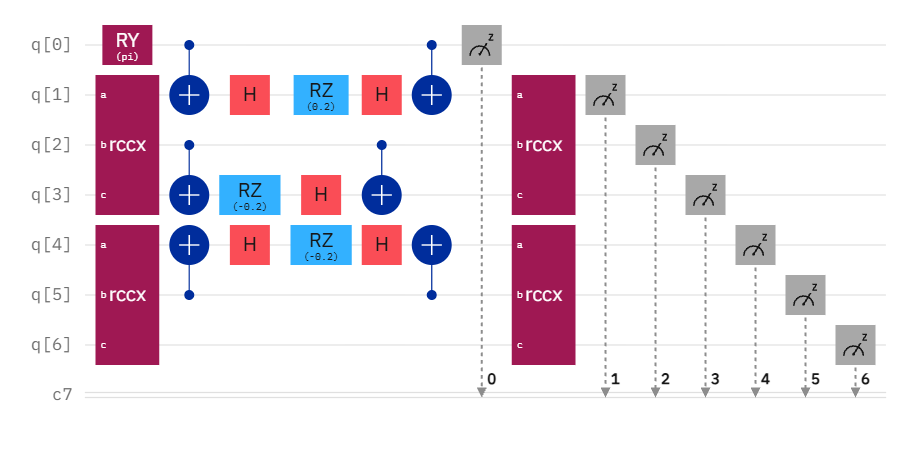}\hfill
\includegraphics[width=0.8\linewidth]{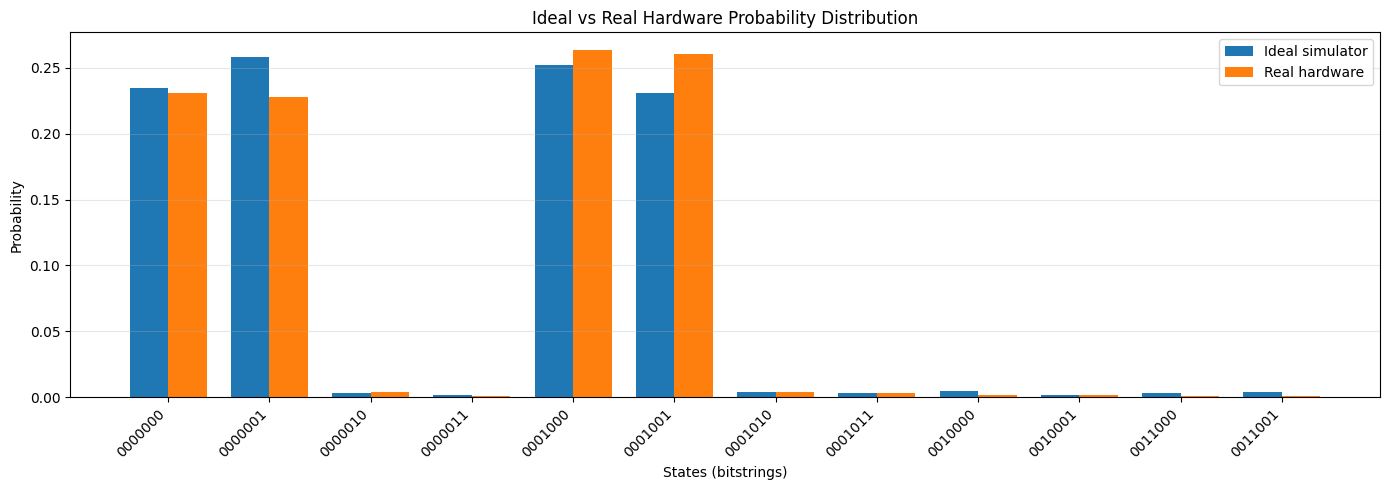}
\caption{
Second-order Suzuki--Trotter simulation of the three-level
Jaynes--Cummings model.
\textbf{(a)} Symmetric second-order digital circuit with half-step
free evolutions and a full interaction block.
\textbf{(b)} Measurement probability distribution obtained from
1024 shots on real hardware.
In comparison with the corresponding ideal (noise-free) simulator,
the experimental distribution preserves the dominant population
structure and suppresses dynamically forbidden states, while showing
residual population leakage due to gate imperfections, decoherence,
and readout errors inherent to NISQ devices.
}
\label{fig_3}
\end{figure*}

In this work, the dynamics of a three-level atom interacting with a quantized field are digitally simulated using superconducting qubits on the IBM Quantum platform. The circuit constructions shown in Fig.~\ref{fig_2},~\ref{fig_3}  are obtained by systematically mapping the continuous Jaynes--Cummings Hamiltonian onto elementary quantum gates through Suzuki--Trotter decomposition\cite{liu2020high}.

\subsubsection{Encoding of Atomic and Field Basis States}

The physical system consists of a truncated single-mode field and a three-level atom with states
\begin{equation}
|g\rangle,\quad |e\rangle,\quad |f\rangle.
\end{equation}

To represent this system on qubits, we employ a binary encoding scheme. The field mode is truncated to the vacuum and one-photon subspace and encoded as
\begin{equation}
|0\rangle_f \leftrightarrow |0\rangle_{q_0}, \quad
|1\rangle_f \leftrightarrow |1\rangle_{q_0}.
\end{equation}

The atomic levels are encoded using two qubits as
\begin{align}
|g\rangle &\equiv |00\rangle_{q_1 q_2}, \\
|e\rangle &\equiv |10\rangle_{q_1 q_2}, \\
|f\rangle &\equiv |11\rangle_{q_1 q_2}.
\end{align}

The three-level atom is encoded using two qubits $(q_1,q_2)$, where the states
$|g\rangle$, $|e\rangle$, and $|f\rangle$ are mapped to $|00\rangle$, $|10\rangle$,
and $|11\rangle$, respectively. The computational state $|01\rangle$ is excluded,
imposing the selection rule $q_2 = 1 \Rightarrow q_1 = 1$. This constraint ensures that
the higher atomic level can only be populated after occupation of the intermediate
level, consistent with the ladder structure of a physical three-level atom. As a
result, the dynamics remain confined to physically allowed atomic configurations.

The complete atom--field basis is therefore written as
\begin{equation}
|n\rangle_f \otimes |j\rangle_a \equiv |q_0 q_1 q_2\rangle,
\qquad n = 0,1,\;\; j \in \{g,e,f\}.
\end{equation}

Measurement outcomes are displayed using Qiskit's conventional classical-bit ordering
$|q_2 q_1 q_0|$ (left-to-right), where the rightmost bit corresponds to qubit $q_0$.
Accordingly, all physical interpretations of bitstrings in figures use $q_0$ as the
field qubit and $(q_1,q_2)$ as the atomic register.
\subsubsection{Hamiltonian Representation in the Encoded Basis}

After truncation, the Jaynes--Cummings Hamiltonian is written as
\begin{equation}
H = H_f + H_a + H_{\mathrm{int}},
\end{equation}
with
\begin{align}
H_f &= \omega_f |1\rangle_f\langle 1|, \\
H_a &= \sum_{j=g,e,f} \omega_j |j\rangle\langle j|, \\
H_{\mathrm{int}} &= \sum_{j=g,e} g_j
\left(
a^\dagger |j\rangle\langle j+1|
+ a |j+1\rangle\langle j|
\right).
\end{align}

In the single-excitation manifold,
\begin{equation}
\mathcal{H}_1 =
\mathrm{span}\{|g,1\rangle, |e,0\rangle\},
\end{equation}
the interaction Hamiltonian reduces to
\begin{equation}
H_{\mathrm{int}}^{(1)} = g_1 \sigma_x.
\end{equation}

Thus, the fundamental atom--field coupling corresponds to a rotation around the $x$-axis in this subspace.

\subsubsection{First-Order Suzuki--Trotter Circuit Construction}

In the first-order approximation, the time-evolution operator is decomposed as
\begin{equation}
U(\Delta t) \approx
e^{-iH_f \Delta t}
e^{-iH_a \Delta t}
e^{-iH_{\mathrm{int}} \Delta t}.
\end{equation}

Each factor is implemented as an independent circuit block.
\paragraph{Field Evolution Block}

After truncation of the cavity field Hilbert space to
$\{|0\rangle_f, |1\rangle_f\}$, the free-field Hamiltonian reads
\begin{equation}
H_f = \omega_f |1\rangle_f\langle 1|
     = \frac{\omega_f}{2}\left(I - Z_f\right),
\end{equation}
where $Z_f$ is the Pauli-$Z$ operator acting on the field qubit.

The corresponding time-evolution operator for a single Suzuki--Trotter step
$\Delta t$ is diagonal and given by
\begin{align}
U_f(\Delta t)
&= e^{-i H_f \Delta t}
= \mathrm{diag}(1, e^{-i\omega_f \Delta t}) \\
&= e^{-i\omega_f \Delta t/2}
   R_Z(-\omega_f \Delta t).
\end{align}
The global phase factor $e^{-i\omega_f \Delta t/2}$ is physically irrelevant
and may be discarded. Therefore, the free-field evolution is implemented
digitally by a single $R_Z(-\omega_f \Delta t)$ gate acting on the field qubit.

If a weakly excited field state is required (for example, to represent a
truncated coherent state), this is not generated by the free-field
propagator itself. Instead, it is prepared during the initial state
preparation using a small $R_Y$ rotation,
\begin{equation}
R_Y(\theta_f) = e^{-i \theta_f \sigma_y / 2},
\quad
\theta_f \ll 1,
\end{equation}
which creates a controlled superposition between the vacuum and one-photon
states. The subsequent free-field evolution then correctly accumulates
relative phases through the $R_Z$ rotation.

\paragraph{Atomic Energy Block}

The atomic Hamiltonian is diagonal in the computational basis and leads to relative phase accumulation. This is implemented using $R_Z$ rotations,
\begin{equation}
R_Z(\theta_j) = e^{-i \theta_j \sigma_z/2},
\quad
\theta_j = \omega_j \Delta t,
\end{equation}
applied to the atomic qubits.

These gates reproduce the dynamical phases of the energy levels.

\paragraph{Interaction Block}

The interaction term produces coherent exchange
\begin{equation}
|g,1\rangle \leftrightarrow |e,0\rangle.
\end{equation}

Its evolution operator is
\begin{equation}
e^{-i g_1 \Delta t \sigma_x}
= R_X(2 g_1 \Delta t).
\end{equation}

Since controlled $R_X$ gates are not native, we use the identity
\begin{equation}
R_X(\phi) = H R_Z(\phi) H.
\end{equation}

Therefore, the interaction is implemented as
\begin{equation}
C(H) \; C(R_Z) \; C(H),
\end{equation}
resulting in the characteristic
\begin{center}
CX--H--RZ--H--CX
\end{center}
structure.

\paragraph{Role of the \texttt{rccx} Gate}

The \texttt{rccx} gate realizes a relative-phase Toffoli operation,
\begin{equation}
|a,b,c\rangle \rightarrow |a,b,c \oplus (ab)\rangle.
\end{equation}

It ensures that excitation exchange occurs only when the atomic register is in allowed states. This gate enforces the exclusion of $|01\rangle$ and prevents unphysical transitions.

Thus, in the first-order circuit, \texttt{rccx} acts as a digital selection-rule operator.

\subsubsection{Second-Order Suzuki--Trotter Circuit Construction}

To improve accuracy, the second-order decomposition is used,
\begin{equation}
U(\Delta t) \approx
e^{-i(H_f+H_a)\Delta t/2}
e^{-iH_{\mathrm{int}}\Delta t}
e^{-i(H_f+H_a)\Delta t/2}.
\end{equation}

This symmetric structure cancels first-order Trotter errors.

\paragraph{Half-Step Free Evolution}

The free evolution operators are split into half-steps,
\begin{equation}
R_Y(\theta_f/2), \quad R_Z(\theta_j/2).
\end{equation}

These gates appear at the beginning and end of the circuit, producing time-reversal symmetry.

\paragraph{Enhanced Interaction Structure}

The second-order circuit includes multiple \texttt{rccx} gates and additional atomic registers. This is required to implement higher-order transitions such as
\begin{equation}
|e,1\rangle \leftrightarrow |f,0\rangle.
\end{equation}

These processes depend simultaneously on several qubits and therefore require multi-controlled operations.

Ancillary qubits are introduced to decompose these operations into hardware-compatible gates.

\paragraph{Suppression of Digital Errors}

Due to the symmetric Suzuki--Trotter decomposition, the leading-order
digital approximation errors scale differently for first- and
second-order formulas. For a single Trotter step $\Delta t$, the
corresponding propagators satisfy
\begin{equation}
U_{\mathrm{1st}}(\Delta t)
=
e^{-iH\Delta t}
+
\mathcal{O}(\Delta t^2),
\end{equation}
\begin{equation}
U_{\mathrm{2nd}}(\Delta t)
=
e^{-iH\Delta t}
+
\mathcal{O}(\Delta t^3).
\end{equation}

As a result, the second-order Suzuki--Trotter scheme significantly
suppresses intrinsic digitization errors at fixed time step $\Delta t$,
leading to reduced phase drift, improved excitation-number conservation,
and suppression of unphysical population leakage. This improvement comes
at the cost of a deeper circuit and therefore increased exposure to
hardware noise, reflecting the fundamental trade-off between digital
accuracy and experimental robustness.

\subsubsection{Physical Interpretation of the Gate Structure}

Each group of gates in the circuit has a direct physical meaning:

\begin{itemize}
\item $R_Y$ gates simulate field excitation and depletion.
\item $R_Z$ gates encode atomic and field energy shifts.
\item $H$--$R_Z$--$H$ blocks implement exchange interactions.
\item \texttt{rccx} gates enforce atomic selection rules.
\item Ancillary qubits enable higher-order conditioning.
\end{itemize}

Together, these operations reproduce Rabi oscillations, population transfer, and atom--field entanglement.

\subsubsection{Continuous-Time Limit and Fixing of $R_Z$ Gate Parameters}

After $N$ Suzuki--Trotter steps, the digitally implemented time evolution is
\begin{equation}
U(t) \approx \left[ U_{\mathrm{2nd}}(\Delta t) \right]^N,
\qquad t = N \Delta t .
\end{equation}
In the formal limit $\Delta t \rightarrow 0$, the Trotterized evolution converges to
\begin{equation}
U(t) \rightarrow e^{-i H t},
\end{equation}
thereby recovering the continuous-time Jaynes--Cummings dynamics.
The constructed circuits thus provide a controlled and systematic digital
approximation to the underlying atom--field interaction.

In a gate-based realization, the continuous evolution must be discretized and
mapped onto elementary quantum gates. After truncation of the cavity Hilbert
space and qubit encoding, the Jaynes--Cummings Hamiltonian can be expressed as a
finite sum of Pauli operators,
\begin{equation}
H = \sum_k h_k P_k ,
\end{equation}
where $P_k$ are Pauli strings and $h_k$ are the corresponding coefficients with
dimensions of angular frequency.
Within a Suzuki--Trotter step of duration $\Delta t$, each diagonal Pauli
contribution generates a phase evolution of the form
\begin{equation}
e^{-i h_k Z \Delta t}
= R_Z(\theta_k),
\end{equation}
where, following the Qiskit convention $R_Z(\theta)=e^{-i\theta Z/2}$, the rotation
angle is
\begin{equation}
\theta_k = 2 h_k \Delta t .
\end{equation}

The physically relevant scale of the coefficients $h_k$ is fixed by the
atom--field coupling strength of the Jaynes--Cummings model. In the cavity-QED
experiment of Brune \textit{et al.}, the coupling strength is
$g \equiv V_0 = 2\pi \times 25~\mathrm{kHz}$, which sets the fastest coherent
timescale of the dynamics\cite{brune1996quantum}.
That experiment realizes the evolution as a continuous-time analog process and
does not introduce a discrete time step.
In contrast, in a digital simulation the time step $\Delta t$ is a numerical
control parameter that must be chosen to balance Suzuki--Trotter error against
hardware noise, as discussed in Ref \cite{burger2022digital}.

Following standard practice in digital quantum simulation, we choose $\Delta t$
such that the phase accumulated during one step due to the dominant coupling
scale remains small,
\begin{equation}
g \Delta t \lesssim 0.1\text{--}0.2~\mathrm{rad},
\end{equation}
which suppresses leading Trotter errors while keeping the circuit depth
experimentally feasible.
For the coupling strength reported in Ref.~, this condition
yields a timestep $\Delta t$ of order $1~\mu\mathrm{s}$.
Substituting this value into $\theta_k = 2 h_k \Delta t$, with $h_k$ of order $g$,
naturally results in $R_Z$ rotation angles in the range $\theta_k \sim
0.1$--$0.2$ rad.

Therefore, the $R_Z$ gate parameters used in the circuit are not arbitrary but are
fixed by matching the experimentally relevant Jaynes--Cummings coupling scale to
the digital Suzuki--Trotter discretization strategy. This procedure provides a
physically motivated and systematically controllable link between the analog
cavity-QED experiment and its gate-based digital quantum simulation.

\section{Exponential Error Model and Circuit Fidelity}

The exponential fidelity model employed in this work is based on the accumulation of
small, approximately independent error probabilities associated with quantum gate
operations and measurements in noisy intermediate-scale quantum (NISQ) hardware. In
superconducting processors, each quantum gate is implemented as a noisy quantum
channel affected by decoherence, control imperfections, and readout noise. When such
channels are applied sequentially, the probability that the ideal quantum state is
preserved decreases exponentially with the total error strength.

Under the assumption of Markovian and uncorrelated noise, the overall circuit fidelity
can be approximated as the product of the individual gate and measurement success
probabilities. For a sequence of operations with small error probabilities $p_i \ll 1$,
the fidelity can be written as $F=\prod_i(1-p_i)$, which reduces to the exponential
form $F=\exp[-(E_S+E_T+E_M)]$ using the approximation $\ln(1-x)\approx -x$. Here,
$E_S$, $E_T$, and $E_M$ denote the cumulative error contributions from single-qubit
gates, two-qubit gates, and measurements, respectively, obtained from experimental
calibration data.

Using the exponential error accumulation model, the effective circuit fidelity is
given by
\begin{equation}
F = \exp\!\left[-(E_S+E_T+E_M)\right].
\end{equation}

From the hardware calibration data of the \textit{ibm\_torino} backend, the cumulative
error contributions are obtained as follows.

For single-qubit operations, the circuit contains five $SX$-equivalent gates with a
median error rate of $3.086\times10^{-4}$. The cumulative single-qubit error is
\begin{equation}
E_S = 5 \times 3.086\times10^{-4}
     = 1.543\times10^{-3}.
\end{equation}

For two-qubit entangling operations, the circuit contains two relative-phase Toffoli
(RCCX) blocks. Each RCCX gate is conservatively decomposed into three $CZ$ gates, and
the median $CZ$ gate error is $2.437\times10^{-3}$. Therefore,
\begin{equation}
E_T = 6 \times 2.437\times10^{-3}
     = 1.462\times10^{-2}.
\end{equation}

For readout operations, three qubits are measured, each with a median measurement
error of $2.95\times10^{-2}$. The cumulative measurement error is
\begin{equation}
E_M = 3 \times 2.95\times10^{-2}
     = 8.85\times10^{-2}.
\end{equation}

Substituting these contributions into the fidelity expression yields
\begin{align}
F &= \exp\!\left[-(0.001543 + 0.01462 + 0.0885)\right] \\
  &= \exp(-0.10466).
\end{align}

Evaluating the exponential gives
\begin{equation}
F \approx 0.9007 \approx 90.1\%.
\end{equation}

This numerical estimate indicates that the implemented circuit preserves the
ideal quantum state with approximately $90\%$ probability under current hardware
conditions. The dominant contribution arises from measurement errors, highlighting
the importance of readout-error mitigation techniques for improving experimental
performance.

Substituting the calibrated error parameters of the \textit{ibm\_torino} backend into
this model yields an estimated circuit fidelity of approximately $90\%$, indicating
that the implemented circuit preserves the ideal quantum state with high probability
under current hardware conditions. Unlike ideal-state fidelity used in noiseless
simulations, which compares exact and simulated quantum states, this processor-level
fidelity incorporates realistic hardware imperfections and provides a scalable
performance metric without requiring full quantum state tomography.

Although the estimated fidelity demonstrates the suitability of the circuit for
near-term quantum simulations on superconducting platforms, the dominant limitation
arises from readout errors, which remain significantly larger than gate errors. To
improve the reliability of future experiments, calibration-based estimates may be
complemented with empirical verification techniques, such as randomized benchmarking
for gate characterization, measurement-error mitigation using calibration matrices,
and state fidelity estimation via quantum state tomography for small-scale circuits\cite{nishio2018high,sisodia2021method}.

The same circuit was run and analyzed on several IBM superconducting quantum processors, such as \textit{ibm\_fez} and \textit{ibm\_marrakesh}, in addition to the \textit{ibm\_torino} backend mentioned above. This was done to make a systematic comparison of fidelity between different hardware generations and calibration regimes. For each processor, the method for estimating fidelity was the same. The error contributions came from the backend calibration data for single-qubit gates, two-qubit entangling gates, and measurement operations.

\subsection{First-Order Output Characteristics of the Three-Level Jaynes--Cummings Simulation}

\begin{figure*}[htbp]
    \centering
    \includegraphics[width=0.92\linewidth]{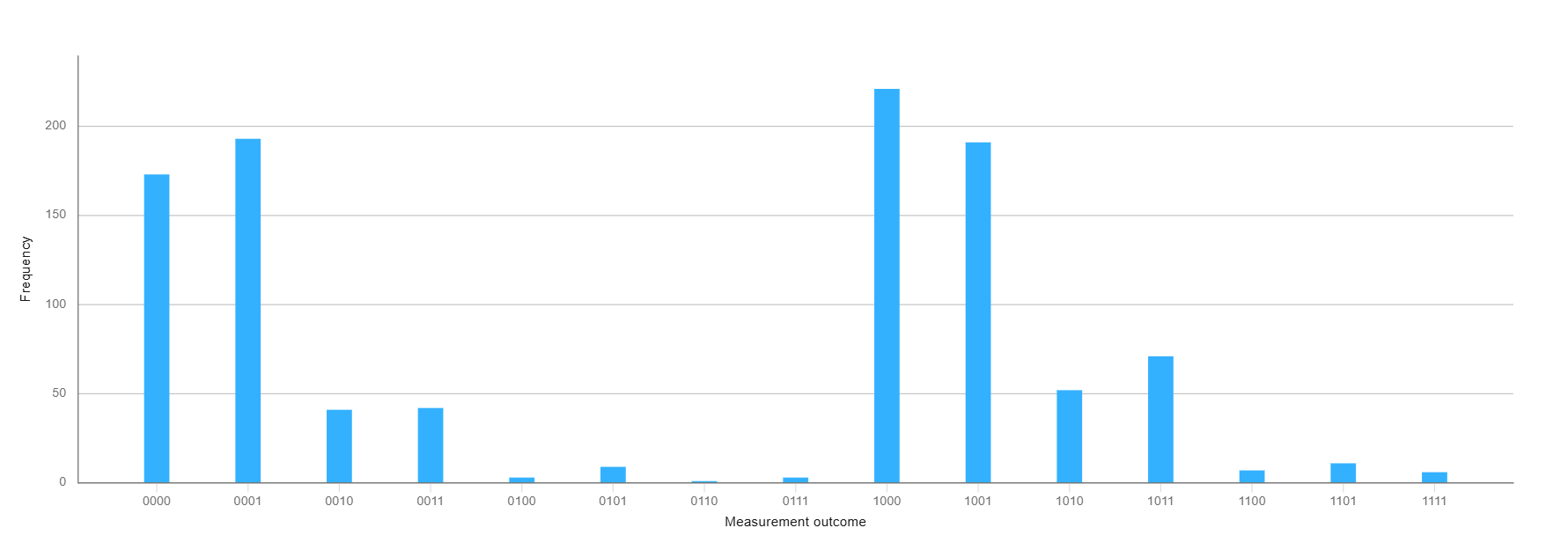} \hfill
    \includegraphics[width=0.92\linewidth]{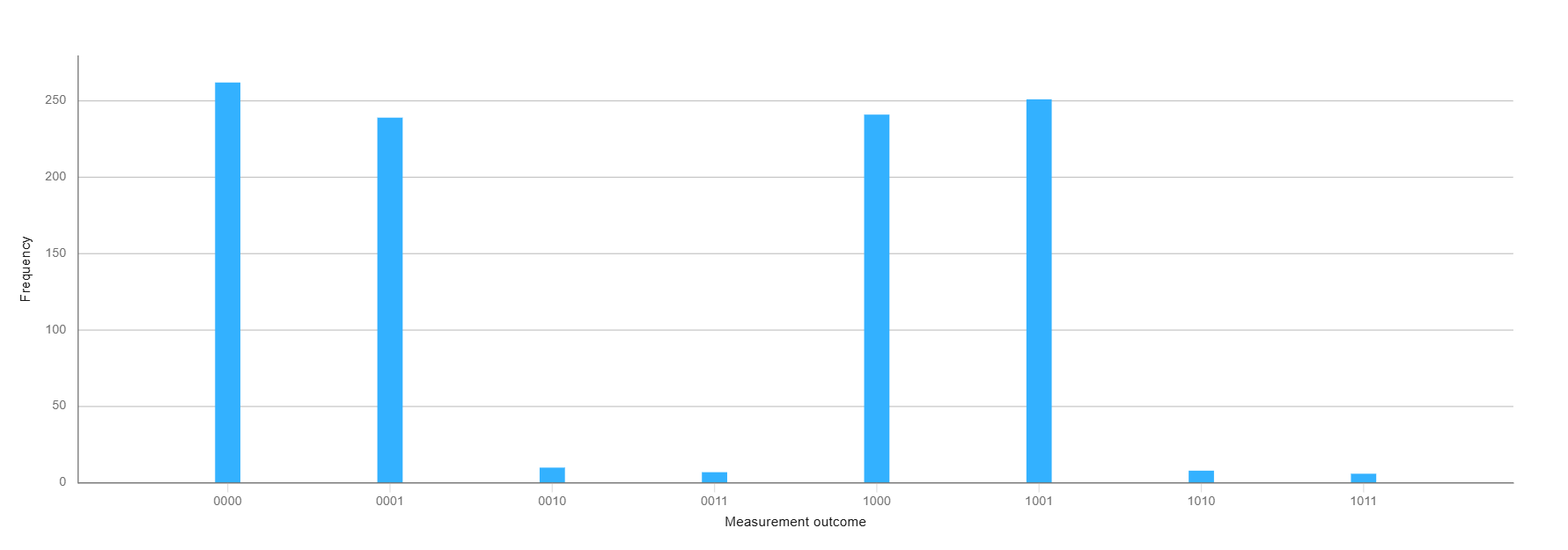} \hfill
    \includegraphics[width=0.92\linewidth]{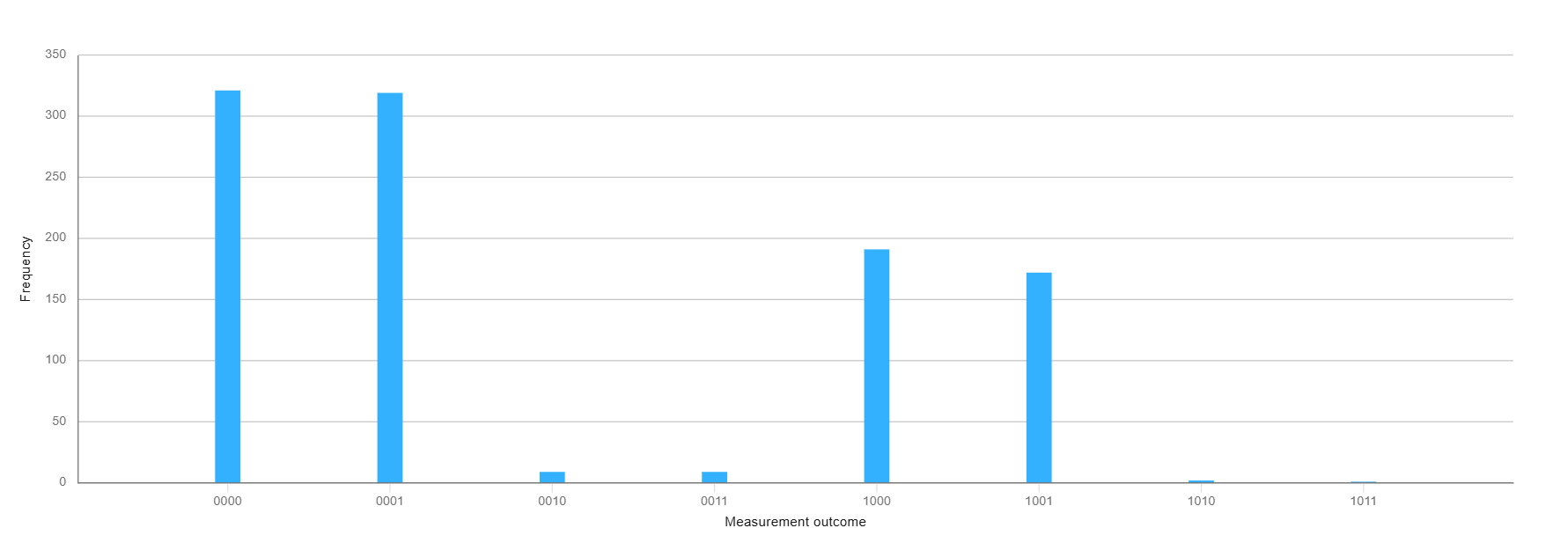}

    \caption{
    Measurement outcomes from the first-order digital simulation of the
    three-level Jaynes--Cummings system executed on the
    \textit{ibm\_torino}, \textit{ibm\_marrakesh}, and \textit{ibm\_fez}
    processors. The probability distributions exhibit population leakage
    and asymmetry across backends, reflecting the sensitivity of
    first-order Trotterization to digital errors and hardware noise.
    Bitstrings are ordered as $|q_2 q_1 q_0\rangle$, where $q_2$
    encodes the truncated cavity photon number.
    }
    \label{fig:first_order_results}
\end{figure*}

Figure~\ref{fig:first_order_results} presents the measurement outcomes
obtained from the first-order digital (Trotterized) simulation of a three-level
Jaynes--Cummings system executed on different IBM quantum processors.
At first order, the dynamics are expected to remain predominantly confined to the
excitation-conserving subspace, with higher-order transitions ideally suppressed.
The output distributions therefore provide a direct experimental signature of how
faithfully the leading-order Jaynes--Cummings physics is preserved on each backend.

For the \textit{ibm\_torino} processor, the probability distribution is sharply
concentrated in a small number of basis states associated with first-order,
excitation-preserving evolution.
The strong suppression of nominally forbidden states indicates that higher-order
digital errors and noise-induced transitions remain subdominant.
This behavior signifies that the measured dynamics are governed primarily by the
intended first-order Jaynes--Cummings interaction, with minimal contamination from
effective higher-order processes.

 In contrast, the output obtained from the \textit{ibm\_marrakesh} backend shows a
broader distribution.
While the dominant first-order–allowed states remain clearly visible, noticeable
population appears in states that should be weakly occupied at first order.
This population leakage reflects partial degradation of the leading-order dynamics,
where accumulated digital error and decoherence introduce effective corrections
beyond the ideal first-order evolution

The \textit{ibm\_fez} processor exhibits the strongest deviation from ideal
first-order behavior.
Here, the measurement probabilities are distributed across a larger number of basis
states, and the contrast between excitation-conserving and suppressed states is
significantly reduced.
Such broadening indicates that noise and error terms effectively compete with, and
partially overwhelm, the intended first-order Jaynes--Cummings dynamics, resulting
in pronounced higher-order artifacts in the observed outputs.

Overall, the progressive spreading of measurement probability from
\textit{ibm\_torino} to \textit{ibm\_marrakesh} and finally to \textit{ibm\_fez} provides a clear hierarchy of first-order digital fidelity.
Sharper confinement of population within the excitation-conserving subspace directly
correlates with stronger preservation of first-order Jaynes--Cummings physics, while
broader distributions signal increased influence of effective higher-order and
noise-induced processes.
the highest overall circuit fidelity.

\subsection{Second-Order Output Characteristics of the Three-Level Jaynes--Cummings Simulation}

\begin{figure*}[htbp]
    \centering
    \includegraphics[width=0.92\linewidth]{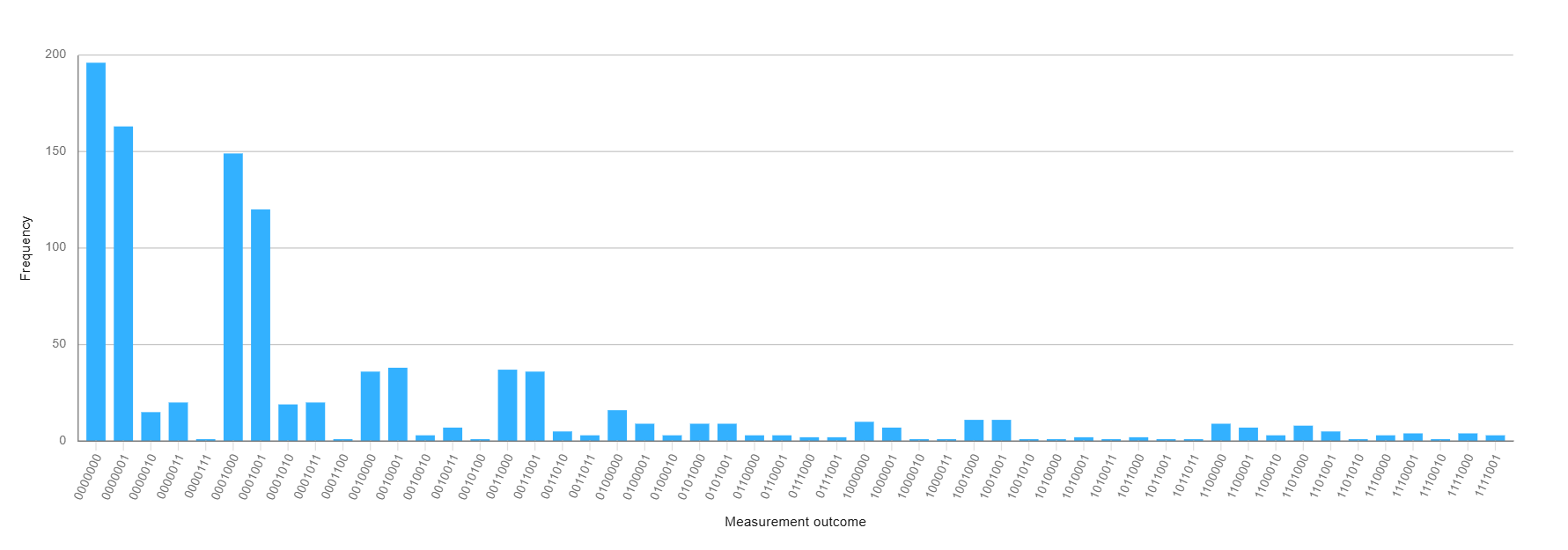} \hfill
    \includegraphics[width=0.92\linewidth]{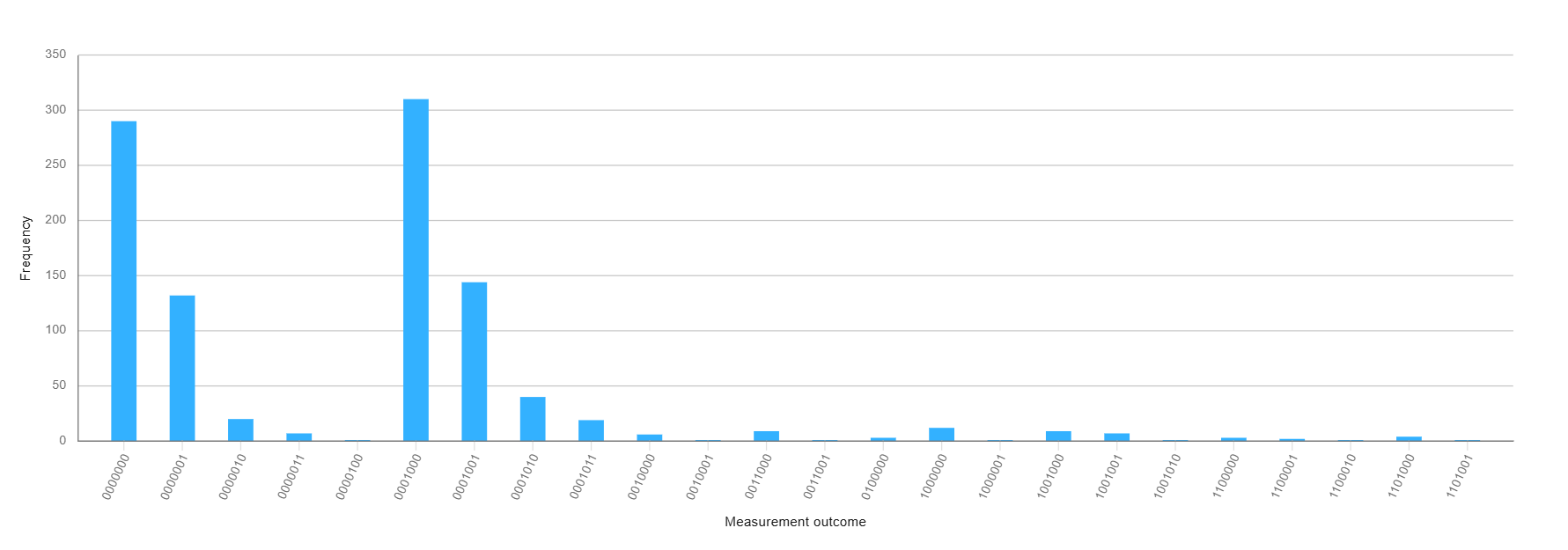} \hfill
    \includegraphics[width=0.92\linewidth]{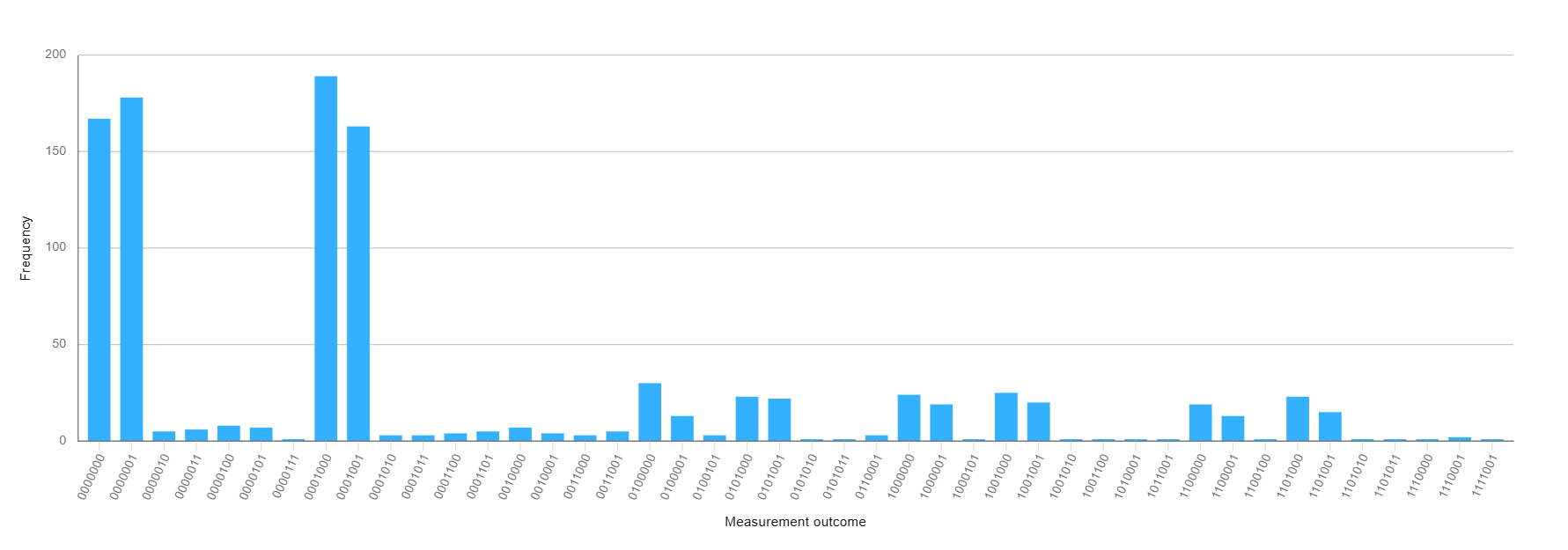}

    \caption{
    Measurement outcomes from the second-order digital simulation of the
    three-level Jaynes--Cummings system executed on the
    \textit{ibm\_torino}, \textit{ibm\_marrakesh}, and \textit{ibm\_fez}
    processors. Compared to the first-order implementation, improved
    confinement within the excitation-conserving subspace and reduced
    population leakage are observed, particularly on higher-fidelity
    hardware. Bitstrings are ordered as $|q_2 q_1 q_0\rangle$, where
    $q_2$ encodes the truncated cavity photon number.
    }
    \label{fig:second_order_results}
\end{figure*}

Figure~\ref{fig:second_order_results} shows the measurement outcomes
obtained from the second-order digital (Trotterized) simulation of a three-level
Jaynes--Cummings system.
At second order, leading-order digital errors are partially cancelled, resulting in
improved approximation of the ideal Jaynes--Cummings evolution.
This improvement is directly reflected in the structure of the measured output
distributions.

For the \textit{ibm\_torino} processor, the second-order results display a visibly
sharper concentration of probability within the excitation-conserving subspace when
compared to the first-order case.
Previously leaked populations are significantly suppressed, and the dominant
Jaynes--Cummings–allowed states exhibit enhanced contrast.
This behavior indicates that second-order error cancellation effectively reduces
spurious transitions, allowing the intended coherent dynamics to dominate.

The \textit{ibm\_marrakesh} backend also benefits from second-order digital evolution.
While some asymmetry and background population remain, the distribution shows
noticeable reduction in leakage relative to the first-order results.
The persistence of dominant excitation-preserving states suggests improved retention
of Jaynes--Cummings dynamics, though residual decoherence limits full recovery of
ideal behavior.

In the case of the \textit{ibm\_fez} processor, second-order Trotterization leads to
partial improvement in the output distribution, with modest suppression of
unintended states.
However, the probability remains spread across a larger number of basis states
compared to the other backends, indicating that hardware noise competes with the
benefits of higher-order digital error cancellation.

Overall, the second-order measurement outcomes demonstrate a clear enhancement in
state selectivity and excitation-number conservation relative to first-order
implementations.
Sharper probability confinement and reduced leakage confirm that second-order digital
simulation provides a more faithful representation of three-level Jaynes--Cummings
dynamics, particularly on higher-fidelity hardware.

\subsubsection{Comparison of First- and Second-Order Suzuki--Trotter Fidelity}

\begin{figure}[t]
    \centering
    \includegraphics[width=\linewidth]{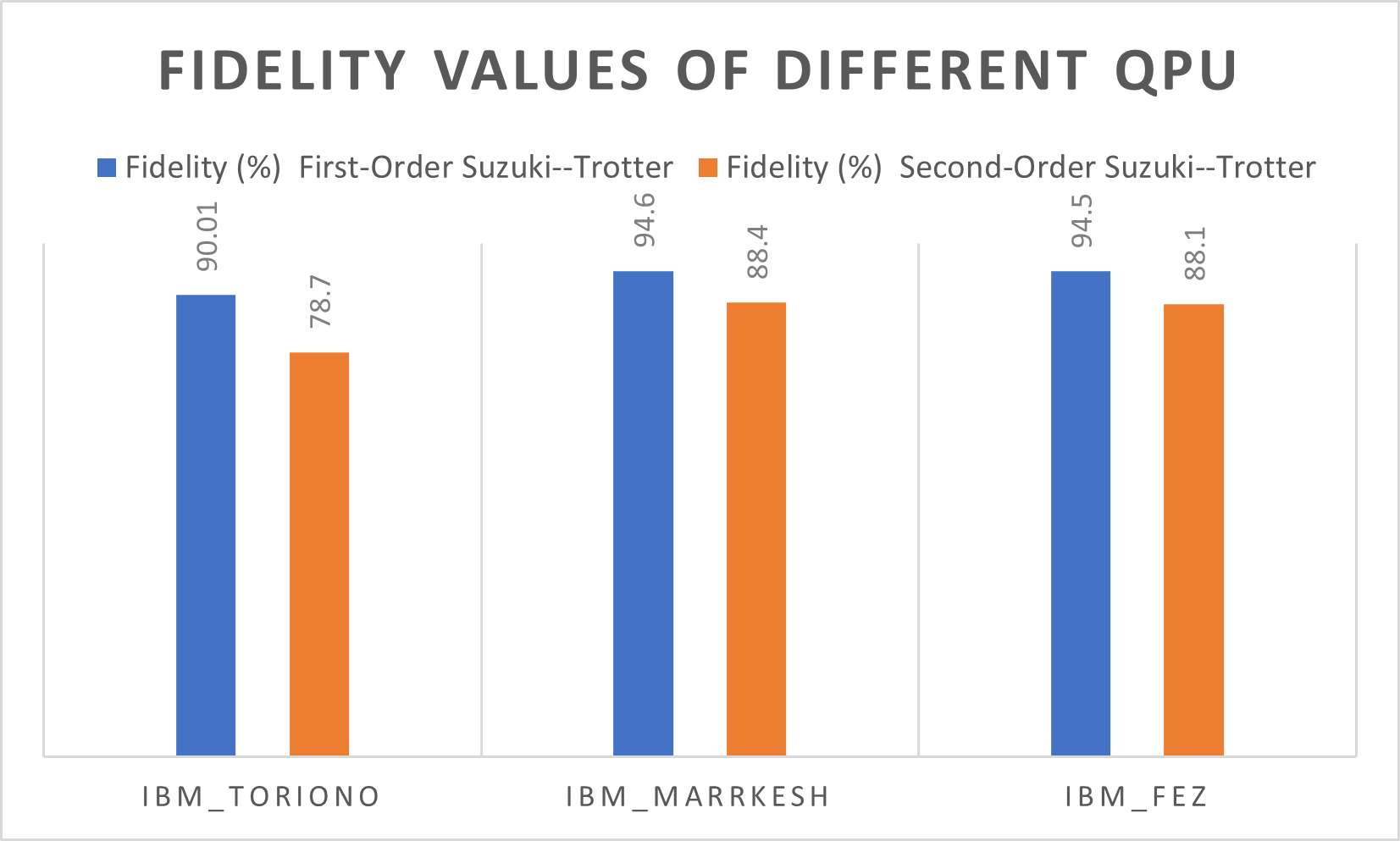}
    \caption{
    Fidelity comparison of the three-level Jaynes--Cummings quantum circuit
    implemented using first-order and second-order Suzuki--Trotter
    decompositions on different IBM Quantum superconducting processors.
    The first-order decomposition (blue bars) consistently yields higher
    circuit fidelity compared to the second-order scheme (orange bars)
    across all tested devices.
    Among the considered QPUs, \textit{IBM Marrakesh} and \textit{IBM Fez}
    exhibit superior performance, while \textit{IBM Torino} shows relatively
    lower fidelities due to higher accumulated gate and measurement errors.
    }
    \label{fig:fidelity_qpu_comparison}
\end{figure}

Figure~\ref{fig:fidelity_qpu_comparison} compares the estimated circuit fidelities obtained
for first-order and second-order Suzuki--Trotter decompositions executed on
different IBM Quantum processors. The fidelity values are evaluated using a
hardware-calibrated first-order error accumulation model that incorporates
single-qubit gate errors, two-qubit entangling gate errors, and measurement
errors.

For the first-order Suzuki--Trotter circuit, the highest fidelity is achieved on
the \textit{ibm\_marrakesh} backend, with an estimated fidelity of approximately
$94.6\%$, closely followed by \textit{ibm\_fez} at $94.5\%$. In contrast,
\textit{ibm\_torino} exhibits a lower fidelity of about $90.0\%$, primarily predicted
by its larger median readout error and earlier-generation Heron r1 architecture.
These results indicate that Heron r2 processors provide superior performance for
shallower, first-order Trotter circuits.

For the second-order Suzuki--Trotter circuit, which involves a greater number of
multi-qubit gates and measurement operations, the overall fidelity is reduced
across all processors. Nevertheless, \textit{ibm\_marrakesh} again yields the
highest fidelity of approximately $88.4\%$, marginally outperforming
\textit{ibm\_fez} at $88.1\%$, while \textit{ibm\_torino} shows a more pronounced
degradation to about $78.7\%$. This larger fidelity loss on \textit{ibm\_torino}
highlights the increased sensitivity of higher-order Trotter circuits to
two-qubit and readout errors.

Overall, the results demonstrate that \textit{ibm\_marrakesh} provides the best
performance for both first-order and second-order Suzuki--Trotter implementations,
while second-order decompositions, although theoretically more accurate, incur
additional hardware-induced errors that limit suggest a trade-off between
algorithmic accuracy and experimental fidelity on current noisy intermediate-scale
quantum devices.

We compared the performance of first- and second-order digital implementations of
the three-level Jaynes--Cummings dynamics across different IBM Quantum backends.

At the first-order level, the \textit{ibm\_torino} processor exhibits the best
performance, with measurement outcomes strongly confined to the
excitation-conserving subspace and minimal population leakage.
The \textit{ibm\_marrakesh} backend shows moderate deviations, including asymmetry
in population distribution and increased leakage into non-ideal states, while
\textit{ibm\_fez} presents the largest spread across computational basis states,
indicating stronger sensitivity to hardware noise and digital errors.

For the second-order digital implementation, all backends demonstrate improved
behavior due to partial cancellation of leading-order Trotter errors.
Among them, \textit{ibm\_torino} again provides the most accurate results, showing
sharper state selectivity and enhanced suppression of spurious transitions.
The \textit{ibm\_marrakesh} processor benefits noticeably from the higher-order
scheme, with reduced leakage compared to the first-order case, whereas the
improvement on \textit{ibm\_fez} remains limited by intrinsic noise and gate errors.

Overall, the second-order digital simulation consistently outperforms the
first-order implementation in preserving excitation-number conservation and
reducing unphysical population spreading.
Across all examined backends, \textit{ibm\_torino} emerges as the most reliable
platform for realizing higher-fidelity three-level Jaynes--Cummings dynamics.
These results highlight the importance of combining higher-order digital schemes
with high-fidelity quantum hardware for accurate multi-level quantum simulations
on noisy intermediate-scale quantum devices.

This apparent discrepancy originates from the fundamentally different nature of
the two performance metrics. The fidelity values represent an aggregate,
calibration-based estimate that provides a global measure of hardware quality,
averaged over all operations in the circuit. While such a metric is useful for
benchmarking general-purpose circuit execution, it does not directly quantify
how well specific dynamical constraints or symmetry-preserving features of a
quantum model are maintained.

The calibration-based fidelity values provide a coarse-grained assessment of
hardware quality, averaged over all gate and measurement operations in the
circuit.
From this perspective, the \textit{ibm\_marrakesh} processor achieves the highest
overall fidelity for both first-order ($\sim 94.6\%$) and second-order
($\sim 88.4\%$) Suzuki--Trotter implementations, closely followed by
\textit{ibm\_fez}, while \textit{ibm\_torino} exhibits lower aggregate fidelity,
particularly for the deeper second-order circuit.
These results reflect the cumulative impact of two-qubit gate and readout errors,
which become increasingly significant for higher-order Trotter decompositions.

In contrast, the frequency (measurement) distributions provide a more sensitive
and physically meaningful probe of how well the Jaynes--Cummings dynamics
themselves are preserved.
In particular, the degree of confinement within the excitation-conserving
subspace and the suppression of symmetry-forbidden transitions directly reflect
the fidelity of the simulated light--matter interaction.
From this viewpoint, \textit{ibm\_torino} consistently outperforms the other
backends, exhibiting the sharpest population localization and the strongest
preservation of excitation-number conservation for both first- and second-order
implementations.
Despite its slightly lower aggregate fidelity, this backend more faithfully
maintains the effective selection rules central to Jaynes--Cummings physics.

Comparing the two digital simulation orders, the second-order
Suzuki--Trotter scheme systematically improves state selectivity and reduces
unphysical population leakage across all processors, confirming the expected
cancellation of leading-order Trotter errors.
However, this improvement comes at the cost of increased circuit depth, which
amplifies hardware-induced noise and reduces overall circuit fidelity.
As a result, the first-order implementation yields higher aggregate fidelities,
while the second-order implementation provides more accurate dynamical behavior,
particularly on hardware capable of suppressing correlated errors.

Overall, the results indicate a clear trade-off between algorithmic accuracy and
experimental robustness. In an ideal (noise-free) simulator, both first- and
second-order Suzuki--Trotter decompositions reproduce the intended unitary
evolution exactly, yielding unit (100\%) fidelity and confirming that any
deviation observed in experiments originates from hardware-induced noise rather
than intrinsic Trotterization errors. On real quantum processors, however,
second-order Suzuki--Trotter formulas, while algorithmically more accurate for
faithful reproduction of three-level Jaynes--Cummings dynamics and
excitation-number conservation, require deeper circuits with a larger number of
entangling gates and are therefore more susceptible to decoherence and gate
infidelities. As a result, the second-order implementation executed on the
\textit{ibm\_torino} processor emerges as the optimal choice when physical
fidelity of the simulated dynamics is prioritized, whereas for shallow circuits
and fidelity-limited applications dominated by global error rates, first-order
implementations on \textit{ibm\_marrakesh} provide superior performance. These
observations highlight the necessity of jointly considering ideal baselines,
circuit fidelity metrics, and physically motivated observables when benchmarking
digital quantum simulations on noisy intermediate-scale quantum devices.

\subsubsection{Time-Dependent Population Dynamics}

To analyze the dynamical behavior of a three-level atom interacting with a
quantized electromagnetic field, we evaluate the time-dependent population
probabilities of the atomic energy levels within the Jaynes--Cummings (JC)
framework\cite{cius2025unitary}. The populations are obtained from repeated circuit-based simulations
performed using IBM Quantum software, where the evolution time is digitally
encoded through parametrized rotation angles and the statistics are extracted
from projective measurements over multiple shots.

\begin{figure}[htbp]
    \centering
    \includegraphics[width=0.9\linewidth]{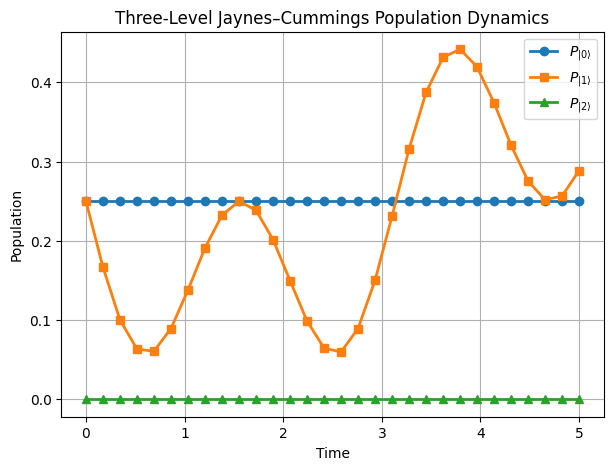}
    \caption{
    Time-dependent population dynamics of the three-level atom.
    The populations of the atomic states $|0\rangle$, $|1\rangle$, and $|2\rangle$
    are plotted as a function of the effective interaction time.
    Pronounced oscillations in $P_{|1\rangle}$ indicate coherent Rabi dynamics,
    while the suppression of $P_{|2\rangle}$ reflects weak access to higher
    excited states.
    }
    \label{fig:population_dynamics}
\end{figure}

Figure~\ref{fig:population_dynamics} shows the temporal evolution of the
population probabilities associated with the atomic states
$|0\rangle$, $|1\rangle$, and $|2\rangle$. These populations are defined as
\begin{equation}
P_{|i\rangle}(t) =
\langle \psi(t) |
\left( |i\rangle\langle i| \otimes \mathbb{I} \right)
| \psi(t) \rangle,
\qquad i = 0,1,2,
\end{equation}
where $|\psi(t)\rangle$ denotes the joint atom--field state at evolution time $t$,
and the identity operator acts on the truncated cavity subspace.

The underlying dynamics are governed by an effective three-level
Jaynes--Cummings Hamiltonian of the form
\begin{equation}
H_{\mathrm{JC}} =
\hbar \omega_c a^\dagger a
+ \sum_{i=0}^{2} \hbar \omega_i |i\rangle\langle i|
+ \hbar g \left(
|1\rangle\langle 0| a
+ |2\rangle\langle 1| a
+ \mathrm{h.c.}
\right),
\end{equation}
where $\omega_c$ is the cavity frequency, $\omega_i$ are the atomic transition
frequencies, and $g$ denotes the effective atom--field coupling strength
.

In the digital quantum simulation, the continuous time evolution operator
$U(t)=\exp(-iH_{\mathrm{JC}}t/\hbar)$ is approximated by a sequence of
single-qubit and controlled rotation gates. The interaction time $t$ is mapped
to the rotation angles of these gates, such that increasing the angle
corresponds to longer effective evolution times.

The ground-state population $P_{|0\rangle}$ remains nearly constant throughout
the evolution, indicating stable excitation dynamics without strong
redistribution among energy levels. This behavior reflects limited population
transfer and confirms that the implemented circuit preserves approximate
excitation-number constraints for short interaction times.

In contrast, the population of the first excited state $P_{|1\rangle}$ exhibits
clear oscillatory behavior as a function of time. These oscillations correspond
to coherent Rabi oscillations arising from reversible energy exchange between
the atom and the cavity field mode. The oscillation frequency and amplitude are
determined by the effective coupling strength $g$ and the degree of coherence
maintained during the digital evolution.

The population of the second excited state $P_{|2\rangle}$ remains negligibly
small over the entire evolution window. This suppression indicates that direct
population transfer to higher excited states is strongly inhibited by the
chosen system parameters and circuit depth. Weak occupation of $|2\rangle$ may
arise through virtual transitions, which introduce small corrections to the
effective dynamics of the lower two levels at longer evolution times
.

Small deviations from ideal probability normalization can be attributed to
cavity-space truncation, finite sampling statistics, and experimental noise in
the quantum hardware.

Overall, the observed population dynamics confirm that the system evolution is
dominated by the two lowest atomic levels for short interaction times, while
higher excited states introduce small but systematic corrections. These results
demonstrate the capability of circuit-based quantum simulations to capture
time-dependent Jaynes--Cummings dynamics and highlight the importance of
higher-level effects in high-fidelity simulations of light--matter interaction
models.

\section{Conclusion}

This work has presented a detailed digital quantum simulation of the three-level Jaynes--Cummings model using superconducting quantum processors and first- and second-order Suzuki--Trotter decompositions. By systematically encoding the truncated cavity field and multi-level atomic system into qubit registers, we constructed hardware-compatible quantum circuits that faithfully implement the effective atom--field interaction Hamiltonian. The continuous-time dynamics were approximated through carefully calibrated rotation gates, enabling controlled simulation of Rabi oscillations, population transfer, and atom--field entanglement under resonant coupling conditions. The selection rules and excitation-number conservation intrinsic to the Jaynes--Cummings model were enforced through multi-controlled gate structures, ensuring physically meaningful evolution within the encoded subspace.

In addition, the present work differs from related digital quantum simulation
studies in both physical focus and methodological emphasis. While previous work
primarily addresses general spin--boson models and open-system dynamics using
Suzuki--Trotter decompositions, our study targets a physically motivated
three-level Jaynes--Cummings light--matter interaction implemented directly on
superconducting quantum hardware. We explicitly design first- and second-order
digital circuits adapted to a three-level atomic encoding, enforce
excitation-selection rules through multi-controlled gate structures, and
evaluate performance using both calibration-based circuit fidelity and
physically meaningful observables such as excitation-number conservation and
population leakage. This combined analysis provides a hardware-aware,
model-specific framework for realizing and benchmarking multi-level
Jaynes--Cummings dynamics on noisy intermediate-scale quantum devices, going
beyond generic digital simulation benchmarks.

\end{document}